\documentclass[aps,pra,notitlepage,twocolumn,superscriptaddress]{revtex4-1}

\usepackage{amsmath,amssymb,amsfonts,amsthm,multirow,graphicx,float,siunitx,dsfont,array}
\usepackage{physics}
\usepackage{graphicx} 
\usepackage[table,xcdraw]{xcolor}
\usepackage[colorlinks,breaklinks]{hyperref}
\usepackage[caption=false]{subfig}
\usepackage{algorithm,algpseudocode}
\newcommand*\diff{\mathop{}\!\mathrm{d}}

\definecolor{darkred}{rgb}{0.5,0,0}
\definecolor{darkgreen}{rgb}{0,0.5,0}
\definecolor{darkblue}{rgb}{0,0,0.5}
\hypersetup{linkcolor=darkred,citecolor=darkgreen,urlcolor=darkblue}
\def\ket#1{|#1\rangle}
\def\braket#1#2{\langle#1|#2\rangle}
\def\ketbra#1{|#1\rangle\langle#1|}

\def\bra#1{\langle#1|}

\newcommand{\nr}{\ensuremath{\hspace*{0.5pt}}}
\renewcommand{\leq}{\leqslant}
\renewcommand{\geq}{\geqslant}

\DeclareMathOperator*{\argmax}{arg\,max}
\DeclareMathOperator*{\argmin}{arg\,min}
\usepackage{tabularx}
\usepackage[framemethod=tikz]{mdframed}
\definecolor{mycolor}{rgb}{0.122, 0.435, 0.698}
\definecolor{mycolor2}{RGB}{166,83,154}
\newmdenv[innerlinewidth=0.5pt,roundcorner=4pt,linecolor=mycolor,innerleftmargin=6pt, innerrightmargin=6pt,innertopmargin=6pt,innerbottommargin=6pt]{mybox}
\usepackage{paracol}
\usepackage{tcolorbox}

\begin{document}
\algnewcommand{\Inputs}[1]{%
  \State \textbf{Inputs:}
  \Statex \hspace*{\algorithmicindent}\parbox[t]{.8\linewidth}{\raggedright #1}
}
\algnewcommand{\Initialize}[1]{%
  \State \textbf{Initialize:}
  \Statex \hspace*{\algorithmicindent}\parbox[t]{.8\linewidth}{\raggedright #1}
}

\title{Programming Quantum Measurements of Time inside a Complex Medium}
\author{Dylan Danese}
\email[Email address: ]{dd52@hw.ac.uk}
\affiliation{Institute of Photonics and Quantum Sciences, Heriot-Watt University, Edinburgh, UK}
\author{Vatshal Srivastav }
\email[Email address: ]{dd52@hw.ac.uk}
\affiliation{Institute of Photonics and Quantum Sciences, Heriot-Watt University, Edinburgh, UK}
\author{Will McCutcheon}
\affiliation{Institute of Photonics and Quantum Sciences, Heriot-Watt University, Edinburgh, UK}
\author{Saroch Leedumrongwatthanakun}
\affiliation{Institute of Photonics and Quantum Sciences, Heriot-Watt University, Edinburgh, UK}
\author{Mehul Malik}
\email[Email address: ]{m.malik@hw.ac.uk}
\affiliation{Institute of Photonics and Quantum Sciences, Heriot-Watt University, Edinburgh, UK}

\date{\today}
\begin{abstract}
{The temporal degree-of-freedom of light is incredibly powerful for modern quantum technologies, enabling large-scale quantum computing architectures and record key-rates in quantum key distribution. However, the generalized measurement of large and complex quantum superpositions of the time-of-arrival of a photon remains a unique experimental challenge. Conventional methods based on unbalanced Franson-type interferometers scale poorly with dimension, requiring multiple cascaded devices and active phase stabilization. In addition, these are limited by construction to a restricted set of phase-only superposition measurements. Here we show how the coupling of spatial and temporal information inside a single multi-mode fiber can be harnessed to program completely generalized measurements for high-dimensional superpositions of photonic time-bin. Using the multi-spectral transmission matrix of the fiber, we find special sets of spatial modes that experience distinct dispersive delays through the fiber. By exciting coherent superpositions of these spatial modes, we engineer the equivalent of large, unbalanced multi-mode interferometers inside the fiber and use them to perform high-quality measurements of arbitrary time-bin superpositions in up to dimension 11. The single fiber functions as a scalable, common-path interferometer for time-bin qudits that significantly eases the experimental overheads of standard approaches based on unbalanced Franson-type interferometers, serving as an essential tool for quantum technologies that harness the temporal properties of light.}

\end{abstract}

\maketitle
\section{Introduction}

The photonic temporal degree-of-freedom has recently emerged as a versatile property of light with powerful applications in quantum technologies. Photons traversing looped fiber architectures have enabled scalable approaches to universal linear optical quantum computing \cite{rohde2015simple}, as well as powerful demonstrations of quantum computational advantage \cite{madsen2022quantum} and multi-photon interference \cite{carosini2024programmable}. In parallel, photons encoded in discrete time-bins have been used to achieve record rates in quantum key distribution \cite{islam2017provably} and noise-resilient entanglement distribution over large-scale distances \cite{bulla2023nonlocal, yu2025quantum}. In this regard, the photonic time-domain acts as a high-dimensional quantum system that allows one to encode $d$-dimensional quantum states or \textit{qudits}. Qudits have attracted significant interest in recent years as they provide a practical route towards high-capacity quantum networks \cite{cozzolino2019high, mirhosseini2015high} and device-independent quantum communication \cite{srivastav2022quick, vertesi2010closing}. In addition, they provide powerful advantages for scalable quantum information processing by reducing circuit complexity and enhancing noise tolerance, with recent demonstrations in platforms ranging from photons \cite{goel2022inverse, lib2025high}, ions \cite{meth2025simulating}, to superconducting circuits \cite{champion2025efficient}.

The generalized measurement of large and complex superpositions of high-dimensional quantum states presents a unique experimental challenge. While measurement techniques for photonic qudits encoded in the transverse-spatial and path-integrated domains have seen significant development in recent years \cite{goel2026quantum, forbes2025progress, wang2020integrated}, scalable methods for measuring large time-bin qudits remain a major research endeavor. The archetypal time-bin qubit measurement utilizes a Franson-type unbalanced interferometer that coherently superposes an early and late time-bin to measure arbitrary time-bin qubit bases \cite{franson1989bell}. Extending this method to $d>2$ dimensions is very difficult as it requires multiple cascaded and actively stabilized interferometers, as well as single-photon detectors with low timing jitter. As a result, time-bin measurements with Franson interferometers have been limited to $d=4$ with bulk optics \cite{ikuta2017implementation} and $d=8$ with integrated components in a recent experiment \cite{yu2025quantum}. These limitations have inspired several alternative approaches for measuring high-dimensional time-bins that leverage random walks in coupled fiber loops \cite{monika2025quantum}, diffraction in time via the temporal Talbot effect \cite{widomski2024efficient}, and dispersion-engineered sum-frequency generation \cite{serino2023realization}. While promising, these techniques come with their own sets of limitations and technological challenges that have limited experimental demonstrations to dimensions $d\leq5$.

\begin{figure}[t] 
  \centering
  \includegraphics[width=\columnwidth]{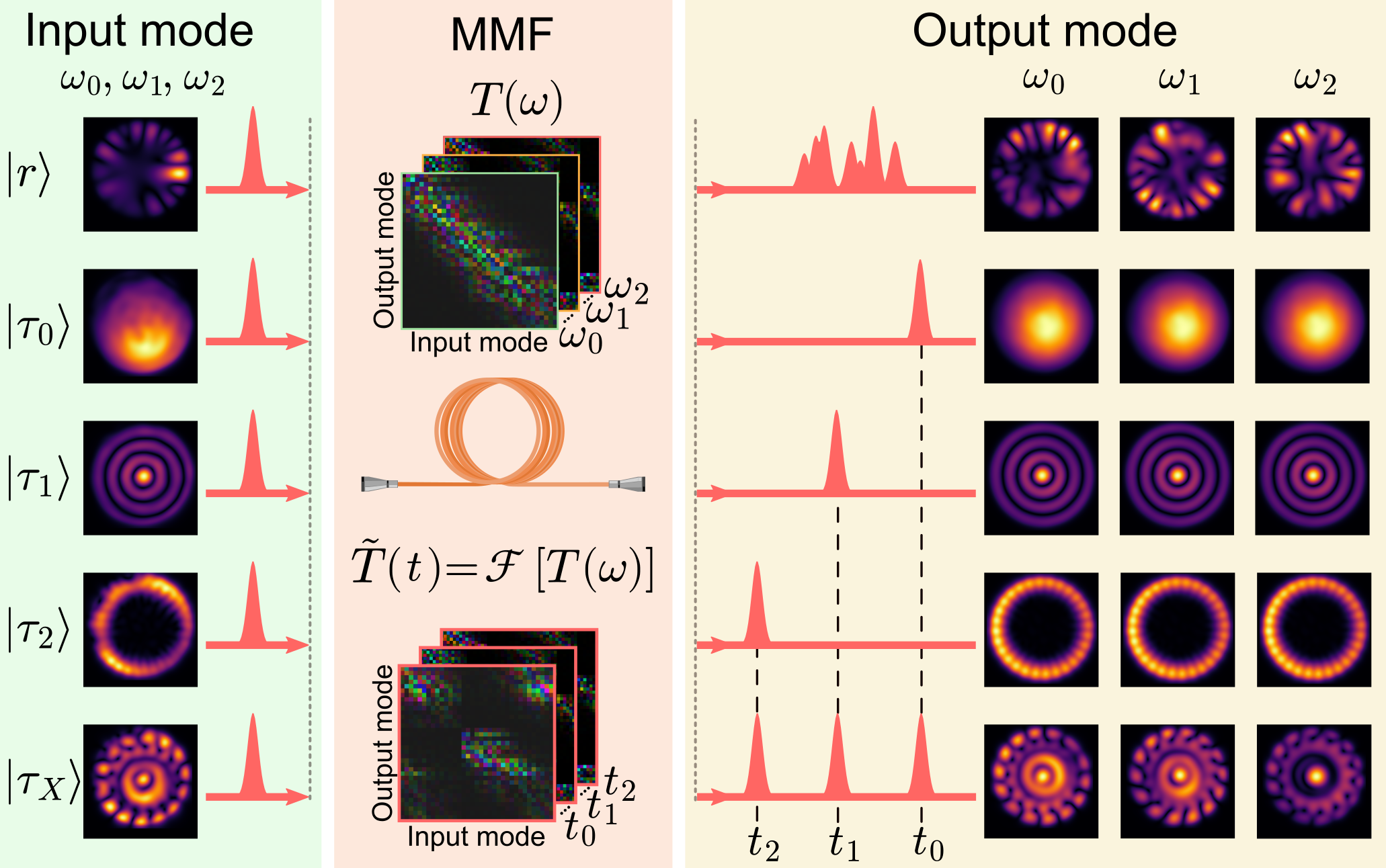}
  \caption{\textbf{$\tau$-modes of a multi-mode fiber (MMF):} A random spatial mode $\ket{r}$ input into an MMF will generally scatter temporally and spectrally. The scattering behavior of an MMF is characterized by its multi-spectral transmission matrix $T(\omega)$, which can be Fourier-transformed to obtain its time-resolved transmission matrix $\tilde{T}(t)$. This matrix can be used to construct a special set of spatial $\tau$-modes ($\ket{\tau_0}$,$\ket{\tau_1}$,$\ket{\tau_2}$) that arrive at distinct times and are spectrally unscattered. A photon prepared in a coherent superposition of these spatial modes $\ket{\tau_X}=\frac{1}{\sqrt{3}}(\ket{\tau_0}+\ket{\tau_1}+\ket{\tau_2})$ would arrive in a non-separable superposition of time and spatial modes $\frac{1}{\sqrt{3}}(\ket{t_0}\ket{\tau_0}+\ket{t_1}\ket{\tau_1}+\ket{t_2}\ket{\tau_2})$. }
  \label{fig:tau_concept}
\end{figure}

Over the past decade, control over light scattering processes inside multi-mode fibers (MMFs) has led to a flurry of exciting advances in classical and quantum photonics \cite{Rotter:2017gb, ploschner2015seeing}. An MMF functions as a large, linear optical network for the photonic spatial, spectral/temporal, and polarization degrees-of-freedom (DoFs), and is characterized by its multi-spectral transmission matrix (MSTM) \cite{andreoli2015deterministic, Mounaix:2015ek}. The ability to experimentally measure the MSTM of an MMF has enabled fundamental studies of principal, super-principal, and anti-principal modes---special states of light that experience strongly enhanced or reduced spectral correlations inside a multi-mode waveguide \cite{carpenter2015observation, Ambichl:2017je}. Alongside this, it has inspired interesting classical photonics applications such as spatial-temporal focusing of ultrashort pulses through an MMF \cite{morales2015delivery, mounaix2019control, mounaix2020time} and fiber-based spectrometers with ultrahigh spectral resolution \cite{redding2014high} . In the quantum realm, control over multi-mode light scattering in the spatial-polarization domains has been used to achieve programmable two-photon interference \cite{defienne2016two, leedumrongwatthanakun2020programmable}, multi-mode entanglement transport \cite{valencia2020unscrambling}, high-dimensional quantum circuits \cite{goel2024inverse}, and reconfigurable multi-user quantum networks \cite{valencia2025large} with a commercial MMF.

\begin{figure*}[t!]
  \centering
  \includegraphics[width=.9\textwidth]{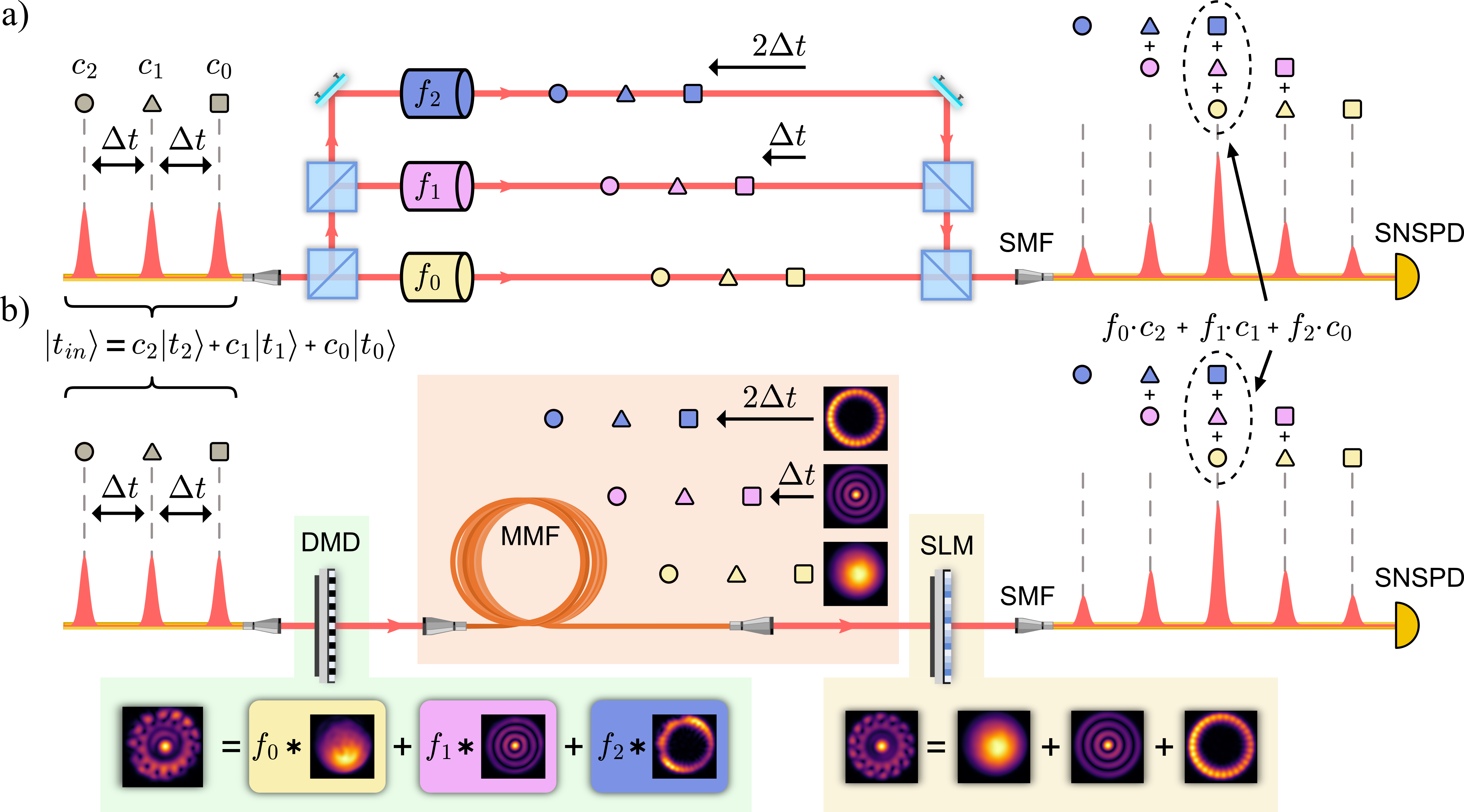}
  \caption{
  \textbf{Programming generalized measurements of time-bin inside a multi-mode fiber (MMF).} \textbf{a)} Schematic showing how a generalized, unbalanced Mach-Zehnder (Franson) interferometer can be used to measure an arbitrary 3-dimensional superposition of time-bins $\ket{t_{in}}$. The shapes (circle, triangle, square) are used as visual aids for labeling each time-bin. $\ket{t_{in}}$ traverses a color-coded, tunable phase-shifter $f_i$ and fixed delay in each arm, resulting in five peaks after the interferometer. The central peak contains a superposition of all three time-bins with desired phase-shifts introduced on each. In this manner, the Franson interferometer acts as a projective measurement of any phase-only time-bin superposition $\ket{\psi} = \frac{1}{\sqrt{3}}(f_0\ket{t_0} + f_1\ket{t_1}+ f_2\ket{t_2})$. Abbreviations: Single-mode fiber, SMF; Superconducting nanowire single-photon detector, SNSPD. \textbf{b)} In our alternative approach, a digital micromirror device (DMD) is used to shape the spatial mode of the time-bin superposition $\ket{t_{in}}$ into an arbitrary superposition of MMF $\tau$-modes depicted in the inset (green). As each $\tau$-mode experiences a different delay in the MMF, the fiber effectively functions as a three-mode unbalanced interferometer as in a), with path replaced by spatial mode. A spatial light modulator (SLM) after the MMF is used to coherently transform the output $\tau$-mode superposition into a Gaussian mode that is coupled into an SMF. The resulting five peaks are the same as those produced in a), except that the MMF approach does not require interferometric stabilization and allows for arbitrary (not just phase-only) superpositions to be measured. In addition, the number of time-bins and their separations can be tuned simply by using an MMF with a different core diameter and length, respectively (see text for more details).}
  
  \label{fig:concept_interferometer}
\end{figure*}

Here we show how the coupling of spatial and temporal information inside an MMF can be harnessed to program generalized measurements for large photonic time-bin qudits. In particular, we use the fiber MSTM to construct a special set of spatial modes that arrive through the MMF unscattered in time. By exciting arbitrary coherent linear superpositions of these modes via wavefront shaping techniques, we are able to engineer large, multi-mode, unbalanced interferometers inside the MMF. In this manner, the MMF functions as a generalized projective measurement in the time-bin basis, allowing us to perform time-bin measurements in mutually unbiased bases (MUBs) in dimensions up to $d=11$ inside a 40m-long commercial step-index MMF. We certify the quality of our measurements by performing quantum tomography via a prepare-and-measure scheme that uses a tuneable Franson interferometer to prepare arbitrary time-bin qubits in all two-dimensional subspaces of our high-dimensional measurement Hilbert space. Our fiber-based method for realizing generalized, high-dimensional measurements of photonic time-bin can readily be scaled to larger time-bin dimensions while not requiring any active phase stabilization, making it particularly exciting for many quantum technology applications. Below, we present the central concept behind our technique, discuss its experimental implementation, and outline its advantages and current limitations.

\section{Concept}

When a random spatial mode $\ket{r}$ localized at a specific time is input into an multi-mode fiber (MMF), the output mode is scattered both temporally and spectrally, i.e.~the output spatial mode changes as a function of the input wavelength (see Fig.~\ref{fig:tau_concept}, top row). The scattering process in the spatial-spectral domains is captured by the fiber multi-spectral transmission matrix (MSTM) $T(\omega)$, which quantifies how a set of input spatial modes map to a set of output spatial modes at a given wavelength $\omega$. The MSTM can be Fourier-transformed to obtain the fiber's time-resolved transmission matrix (TRTM) $\tilde{T}(t)$, which captures the spatial scattering behavior at a given time $t$. The TRTM can be used to construct a special set of orthogonal spatial modes that we call $\tau$-modes $\ket{\tau_i}$. The $\tau$-modes are obtained via a singular value decomposition (SVD) of the TRTM, and have been demonstrated experimentally for controllable energy deposition with ultrashort pulses \cite{devaud2022temporal}. As seen in Fig.~\ref{fig:tau_concept}, the $\tau$-modes ($\ket{\tau_0}$,$\ket{\tau_1}$,$\ket{\tau_2}$) have the property that they arrive temporally and spectrally unscattered. However, the arrival time and spatial structure of each mode is different and is a function of the complex spatial-mode coupling and dispersion occurring within the MMF, characterized by the TRTM. A $\tau$-mode arriving at time $t$ is obtained from the singular vector corresponding to the largest singular value of the TRTM SVD at a given time $t$. The $\tau$-modes are equivalent to the set of principal modes under the assumption of first-order dispersion in the MMF (see Supplementary Section \ref{sec:tau_principal_modes} for details on $\tau$-mode construction and comparison with principal modes). More importantly, due to linearity, the set of $\tau$-modes can be coherently combined to create new bases of modes. Fig.~\ref{fig:tau_concept} depicts an example of such a superposition mode $\ket{\tau_X}=\frac{1}{\sqrt{3}}(\ket{\tau_0}+\ket{\tau_1}+\ket{\tau_2})$ that would arrive in a non-separable superposition of temporal and spatial modes $\frac{1}{\sqrt{3}}(\ket{t_0}\ket{\tau_0}+\ket{t_1}\ket{\tau_1}+\ket{t_2}\ket{\tau_2})$ after traveling through an MMF.

Figure \ref{fig:concept_interferometer}a depicts a standard Franson interferometer modified for measuring arbitrary superpositions of three time-bins given by $\ket{t_\textrm{in}}=c_0\ket{t_0}+c_1\ket{t_1}+c_2\ket{t_2}$. The interferometer consists of three unbalanced paths with a fixed delay corresponding to the time difference $\Delta t$ between two adjacent time-bins. Each path additionally contains a tuneable phase-shifter $f_i$ that introduces a phase $e^{i\phi_i}$. After traversing the interferometer, the time-bin state $\ket{t_\textrm{in}}$ is split into three delayed copies, which are recombined to produce five time-delayed peaks. The central peak results from the interference of all three input time-bins, and its amplitude is proportional to the dot product of individual phase-shifter and time-bin probability amplitudes given by $f_0.c_0+f_1.c_1+f_2.c_2$. By setting the three phase-shifters to specific values $f_i$, the interferometer can be used to perform projective measurements of any phase-only time-bin superposition state given by $\ketbra{\psi}$, where $\ket{\psi} = \frac{1}{\sqrt{3}}(f_0\ket{t_0} + f_1\ket{t_1}+ f_2\ket{t_2})$. Generalizing this device to $d$ dimensions requires $d$ unbalanced paths, resulting in $2d+1$ interference peaks. While here we have shown a multi-path interferometer for simplicity, in practice this is achieved by cascading multiple two-path interferometers \cite{ikuta2017implementation, yu2025quantum}. Such implementations carry a significant overhead in terms of experimental complexity and stability, which has limited current demonstrations to $d\leq8$. In addition, the cascaded approach does not give access to arbitrary $d$-dimensional superposition measurements.

Figure \ref{fig:concept_interferometer}b depicts the central concept of our alternative approach. A time-bin superposition state $\ket{t_\textrm{in}}$ in a Gaussian spatial mode is incident on a digital micromirror device (DMD) that allows us to transform it into an arbitrary spatial mode before it is input into an MMF. We can choose this spatial mode to be an individual $\tau$-mode of the MMF, which would introduce a specific delay onto $\ket{t_\textrm{in}}$ (as shown in Fig.~\ref{fig:tau_concept}). By judiciously choosing this spatial mode to be in a coherent superposition $\ket{\chi} = \frac{1}{\sqrt{3}}(f_0\ket{\tau_0} + f_1\ket{\tau_1}+ f_2\ket{\tau_2})$ of three MMF $\tau$-modes, we can engineer the equivalent of a three-path unbalanced interferometer inside the MMF. A key difference from the bulk-optics approach in Fig.~\ref{fig:concept_interferometer}a is that here, the $\tau$-mode probability amplitudes $f_i$ can be complex, allowing us to implement tunable unbalanced interferometers that perform arbitrary (amplitude and phase) projective measurements of time-bin superposition states. Note that the delays introduced by the $\tau$-modes must be equivalent to the time-bin separations $\Delta t$, which ensures that the different copies interfere effectively. A spatial light modulator (SLM) at the output of the MMF is used to coherently transform the output $\tau$-mode superposition back into a fundamental Gaussian mode that is coupled into a single-mode fibre (SMF), leaving only the temporal interference information to be detected by a time-resolving detector (SNSPD). 

The DMD-MMF-SLM setup is hence functionally equivalent to the multi-path Franson interferometer but utilizes the dispersive nature of the $\tau$-modes instead, giving us access to long delay times. The common-path interferometer implemented in this manner is also phase-stable and can readily be scaled to higher dimensions via an appropriate choice of fiber diameter and length. We experimentally realize this scheme with a 40-m-long commercial step-index MMF to implement programmable time-bin measurements in dimensions $d\in\{2,4,11\}$. First, we measure the MSTM of the MMF in the circularly polarized (CP) spatial-mode basis \cite{ploschner2015seeing,snyder1978modes} by using a digital micromirror device (DMD) to generate 125 CP modes with a narrow-band swept laser source at $1550$ $nm$. These are launched into the MMF and the output complex field amplitude is reconstructed with an InGaAs camera using off-axis holography. We repeat this process for 421 wavelength steps of $3.8$ $pm$ each, resulting in an MSTM with $421\times125\times125$ complex values. The measured MSTM is then Fourier-transformed to obtain a time-resolved transmission matrix (TRTM) with a resolution of $5$ $ps$ over a temporal range of $\approx1.6$ $ns$. Performing the SVD procedure on this TRTM results in 70 $\tau$-modes with a resolvable amplitude. From these, we choose 11 $\tau$-modes with a time-separation of $\approx160$ $ps$ to act as our time-bin basis (see Supplementary section \ref{sec:MSTM_measurement} for additional details).

\section{Experimental Results}

\begin{figure}[t] \includegraphics[width=1\columnwidth, scale=0.4]{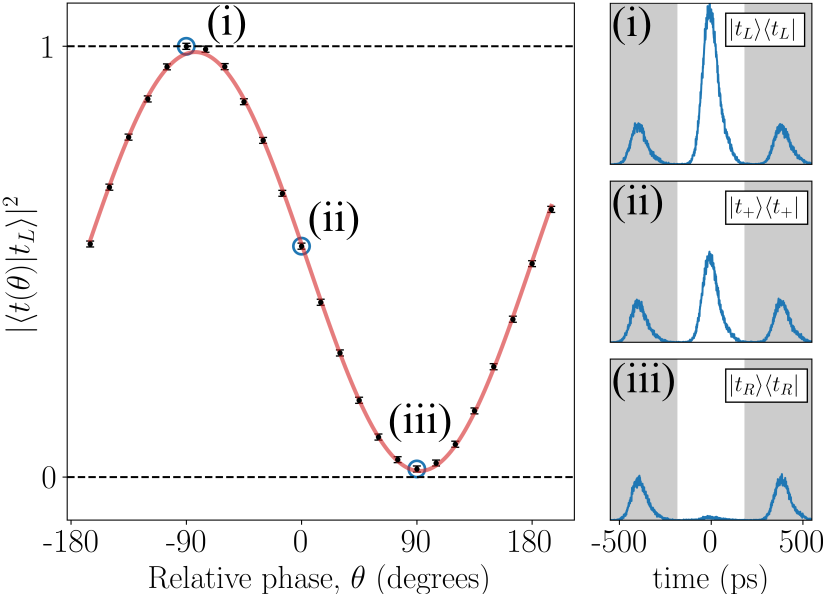}
  \caption{\textbf{Malus' law for a time-bin qubit.} A sinusoidal intensity pattern is observed when we prepare a time-bin qubit $\ket{{t}_{L}}=\frac{1}{\sqrt{2}}(\ket{t_0}-i\ket{t_1})$ and implement a tunable measurement $\ketbra{t(\theta)}$ on it. The phase $\theta$ is varied by changing the $\tau$-mode superposition hologram on the DMD (see Fig.~\ref{fig:concept_interferometer}b). Sets of three interference peaks measured at the output when implementing time-bin measurements (i) $\ketbra{t_L}$, (ii) $\ketbra{t_+}$, and (iii) $\ketbra{t_R}$ in our measurement device. The central peak corresponds to the probability of the measured time-bin state.}
  \label{fig:time_bin_polarizer}
\end{figure}

We test the operation of our device as a programmable time-bin measurement by first using it to perform a tunable projective measurement of a time-bin qubit $\ket{{t}_{L}}=\frac{1}{\sqrt{2}}(\ket{t_0}-i\ket{t_1})$ prepared with a standard, two-mode Franson interferometer. By setting the DMD/SLM to prepare/measure specific input/output $\tau$-mode superpositions in $d=2$, we program our device to perform a tunable, time-bin qubit measurement $\ketbra{t(\theta)}$, where $\ket{t(\theta)}=\frac{1}{\sqrt{2}}(\ket{t_0}+e^{i\theta}\ket{t_1})$. In $d=2$, we expect our device to produce three peaks, with the counts contained in the central peak corresponding to the desired measurement. The resulting count rate as a function of relative phase $\theta$ is shown in Fig.~\ref{fig:time_bin_polarizer}. The visibility curve seen here depicts Malus' law for polarization \cite{malus1809propriete} applied to time-bin qubits. The minimum and maximum points on the curve indicate projective measurements of the orthogonal time-bin states $\ket{{t}_{R,L}}=\frac{1}{\sqrt{2}}(\ket{t_0}\pm i\ket{t_1})$ with $\theta\in\{\pi/2,3\pi/2\}$, while the middle of the curve corresponds to measurements of the mutually unbiased basis states $\ket{{t}_{D,A}}=\frac{1}{\sqrt{2}}(\ket{t_0}\pm \ket{t_1})$ with $\theta\in\{0,\pi\}$. The visibility of cosine-squared fit to the data in Fig.~\ref{fig:time_bin_polarizer} is $0.97$, which shows the high extinction ratio of our time-bin ``polarizer.'' 

Next, we test the operation of our measurement device by implementing projective measurements of high-dimensional ($d\geq 2$) time-bin superposition bases known as mutually unbiased bases (MUBs). MUBs are of significant fundamental interest in quantum information science \cite{Tavakoli_2021_MUBBell} and find many useful applications such as efficient entanglement witnesses and quantum state tomography \cite{Bavaresco2017MeasurementsIT, Guta_2020}. In prime-power dimensions $d$, one can construct $d+1$ possible mutually unbiased bases (MUBs), with one possible construction given by \cite{Wootters:1989ba}:
\begin{align}
    \ket{{v}^\mu_{a}} &= \tfrac{1}{\sqrt{d}} \sum_{t=0}^{d-1} \omega^{at+\mu t^{2}} \ket{t}.
    \label{eq:mub}
\end{align}
Here, $a$ and $\mu$ label the state and basis respectively, $\omega=\exp(\tfrac{2\pi \nr i}{d})$ are the complex $d$-th roots of unity, and $\ket{t}$ are states in the time-bin (computational) basis. States from two different MUBs have an equal overlap of $1/d$ with each other.

\begin{figure*}[t!]  \includegraphics[width=\textwidth]{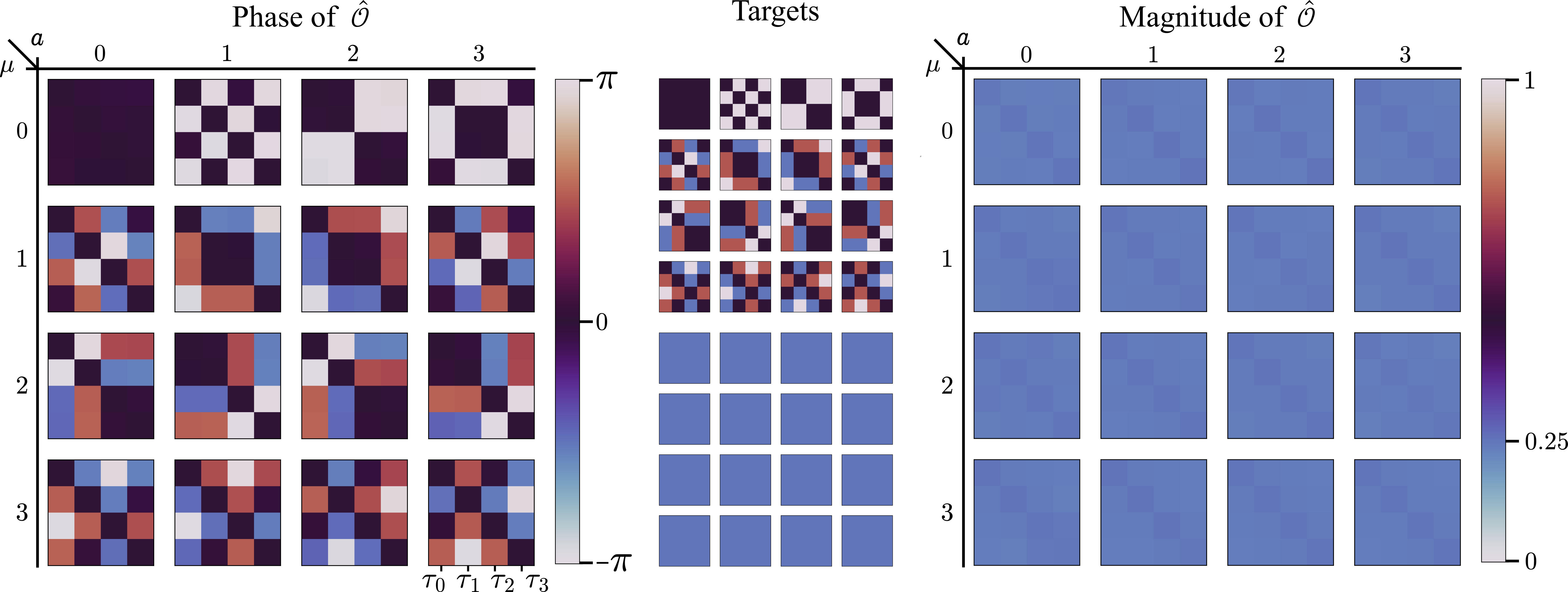}
  \caption{\textbf{Tomography of four-dimensional time-bin measurements.} The left and right-most plots depict the phases and absolute values of 16 tomographically reconstructed measurement operators, $\hat{\mathcal{O}}$. These operators target the 16 possible MUB projectors for 4-dimensional time-bin superpositions $\ket{{v}^\mu_{a}}\bra{{v}^\mu_{a}}$, which are shown in the middle for comparison. We observe an average fidelity of $96.24\%$ between the reconstructed measurement operators and their corresponding target projectors.}
  \label{fig:4dimtomography}
\end{figure*}

We perform projective measurements of these $d$-dimensional time-bin MUB states $\ket{{v}^\mu_{a}}$ by setting our device to display spatial MUBs of the corresponding input/output $\tau$-modes. Once the time-bin measurement is programmed, we characterize it using quantum tomography \cite{Hradil1997,qi2013,Smolin2012,flammia2012}. This involves probing the measurement device with a set of informationally complete time-bin states in $d=2$, and tomographically inverting the resulting data to obtain a density matrix that corresponds to the implemented measurement operator $\hat{\mathcal{O}}$. We can then calculate its fidelity to the target measurement $F=\bra{{v}^\mu_{a}}\hat{\mathcal{O}}\ket{{v}^\mu_{a}}$. We use a tunable two-mode Franson interferometer to generate time-bin superpositions in every two-dimensional subspace of our $d$-dimensional space and input them into the measurement device. The Franson interferometer is phase-locked for stability and includes a $30$-$cm$-long motorized stage in one arm, allowing us to access time-delays across the full $1.6$ $ns$ range (see Supplementary sections \ref{sec:SuppTomo} and \ref{sec:SuppFranson} for additional details on quantum tomography and the tunable Franson respectively). 

 We implement time-bin MUB measurements in dimensions $d\in\{2,4,11\}$. The reconstructed density matrices corresponding to all 16 time-bin MUB measurement operators in dimension $d=4$ are shown in Fig.~\ref{fig:4dimtomography}. We don't measure the computational (time-bin) basis as that is trivially obtained by using a time-resolving detector. The figure depicts the phase and magnitude of the reconstructed density matrix elements, alongside the target density matrices for comparison. As an example, the first row shows the four measurement operators corresponding to the target states in the first MUB $(\mu=1)$:
\begin{align}
\ket{{v}^1_{1}}&=(\ket{t_0}+\ket{t_1}+\ket{t_2}+\ket{t_3})/2\nonumber\\
\ket{{v}^1_{2}}&=(\ket{t_0}-\ket{t_1}+\ket{t_2}-\ket{t_3})/2\nonumber\\
\ket{{v}^1_{3}}&=(\ket{t_0}+\ket{t_1}-\ket{t_2}-\ket{t_3})/2\nonumber\\
\ket{{v}^1_{4}}&=(\ket{t_0}-\ket{t_1}-\ket{t_2}+\ket{t_3})/2.
\end{align}

\begin{figure*}[t!] \includegraphics[width=\textwidth]{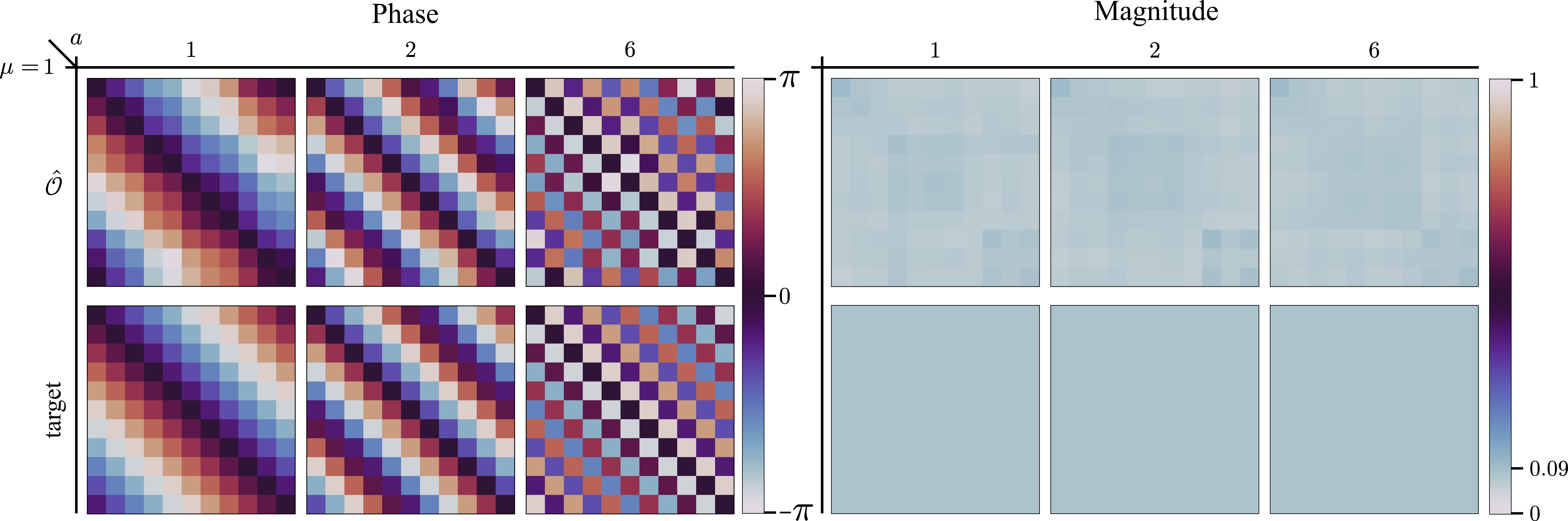}
  \caption{\textbf{Tomography of eleven-dimensional time-bin measurements.} The top row shows the angle and absolute values of three reconstructed time-bin MUB measurement operators $\hat{\mathcal{O}}$ in dimension $d=11$ ($\ketbra{v^{1}_1}$, $\ketbra{v^{1}_2}$ and $\ketbra{v^{1}_6}$). Their corresponding target projectors, shown in the bottom row, are seen to agree qualitatively with the reconstructed measurement operators. We calculate a fidelity of $84.2\%$, $86.3\%$, and $83.7\%$ to the target measurements with $a=1$,2, and 6 respectively.}  \label{fig:11dimtomo}
\end{figure*}

Figure~\ref{fig:11dimtomo} (top row) shows reconstructed density matrices corresponding to 3 of the 132 possible time-bin MUB measurement operators in dimension $d=11$. These correspond to time-bin state superpositions in the first MUB, given by Eq.~\eqref{eq:mub} with $\mu=1$ and $a\in\{1,2,6\}$. The target density matrices are shown in the bottom row. As can be seen in Figs.~\ref{fig:4dimtomography} and \ref{fig:11dimtomo}, the measured and target density matrices show good qualitative agreement with each other. To further quantify them, we calculate their fidelity to their target measurement operators. The average fidelities of the reconstructed time-bin measurement operators in each dimension is shown in Table \ref{tab:avg_fid_table}. The measurement fidelity is seen to drop as dimension is increased. We attribute this to errors introduced in the tomography procedure, which involves the preparation of a large number of time-bin qubit probe states and probing the device over a long period of time. This introduces instabilities in the MSTM due to temperature fluctuations, which reduces the quality of the measurements as the dimension is increased.

\begin{table}[b!]
\caption{Average fidelities of reconstructed measurement operators in dimensions $d\in\{2,4,11\}$ to the ideal target measurements. The fidelity of each individual measurement can be found in the supplementary table \ref{tab:S_fid_all_dims}. Errors are reported to one standard deviation.}
\label{tab:avg_fid_table}
\begin{ruledtabular}
\begin{tabular}{lccc}
Dimension & 2 & 4 & 11 \\
\hline
Fidelity & 98.15 $\pm$ 0.39\%  & 96.24 $\pm$ 0.30\%  & 84.73 $\pm$ 1.27\%  \\
\end{tabular}
\end{ruledtabular}
\end{table}

\section{Discussion and Outlook}

We have proposed and demonstrated a method for measuring an arbitrary, high-dimensional time-bin superposition state by using the natural scattering processes inside a commercial multi-mode fiber (MMF). By precisely shaping an incoming time-bin superposition in its spatial structure, we can control its propagation in time through the MMF. This allows us to engineer the equivalent of large, multi-mode unbalanced interferometers inside the fiber, enabling its use as a programmable measurement device for arbitrary time-bin superposition states. We demonstrate the functionality of our device as a tunable projective measurement for mutually unbiased time-bin qudit states in up to dimension $d=11$ and reconstruct them via quantum tomography.

Our use of a single fiber is significant as it provides a simple pathway towards scaling such measurements to very large dimensions---a larger core diameter would support more spatial, and consequently, more time-bin modes and a longer fiber would ease timing jitter requirements. Above all else, the single fiber serves as a stable, common-path interferometer with large delays, surpassing the stringent alignment and stability requirements of bulk optical setups and the limited temporal delays possible on integrated platforms \cite{ikuta2017implementation,yu2025quantum}. Interestingly, the standard approach based on cascading $N$ two-mode interferometers with $N$ independent phase controls for measuring $d=2^N$ dimensional states cannot access general transformations in dimension $d$. In contrast, our technique is completely general as it allows independent control over each time-bin amplitude.

Our technique currently suffers from a reduction in measurement quality as dimension is increased. While this appears to be a result of the temperature fluctuations over the time period required to tomographically reconstruct our measurement operators, it can be mitigated by incorporating analytic models for how small changes in temperature affect modal dispersion in the MMF \cite{redding2014high} and using an electro-optic modulator to prepare probe states in a fast and stable manner. In addition, the loss introduced during mode generation by the DMD due to its binary modulation can be reduced significantly by using a more efficient device such as a fast SLM or MEMS phase-only light modulator \cite{rocha2024fast}, as well as by incorporating more efficient mode conversion techniques \cite{Hiekkamaki:19}. Our method serves as an exciting fundamental advance for quantum measurement science, where the measurement of large superpositions of time has been a long-standing challenge. It also holds significant potential for the measurement of high-dimensional time-bin and energy-time entanglement \cite{xavier2025energy, jha:2008}, making it particularly exciting for applications in long distance quantum networks with strong noise \cite{srivastav2022quick,bulla2023nonlocal}. Finally, it could find interesting applications in photonic quantum computing, where time-based architectures have recently attracted a lot of interest \cite{Bouchard2024,Humphreys2013,takeda2019toward}.

\vspace{5pt}
\begin{acknowledgements}
  We would like to acknowledge David Phillips for helpful discussions and extend our thanks to Joel Carpenter for his library ``digHolo'' for performing high-speed off-axis holography. We acknowledge financial support from the UK Engineering and Physical Sciences Research Council (EPSRC) (EP/Z533208/1) (EP/Z533166/1), European Research Council (ERC) Starting Grant PIQUaNT (950402), and the Royal Academy of Engineering Chair in Emerging Technologies programme (CiET-2223-112).
\end{acknowledgements}

\bibliography{main.bib}

\newpage
\onecolumngrid

\section{Methods}

\subsection{Multi-spectral transmission matrix (MSTM) measurement} \label{sec:MSTM_measurement}
\subsubsection{Complex field generation}
\label{sec:complex_field_generation}
\begin{figure}[h]
\includegraphics[width=1\columnwidth]{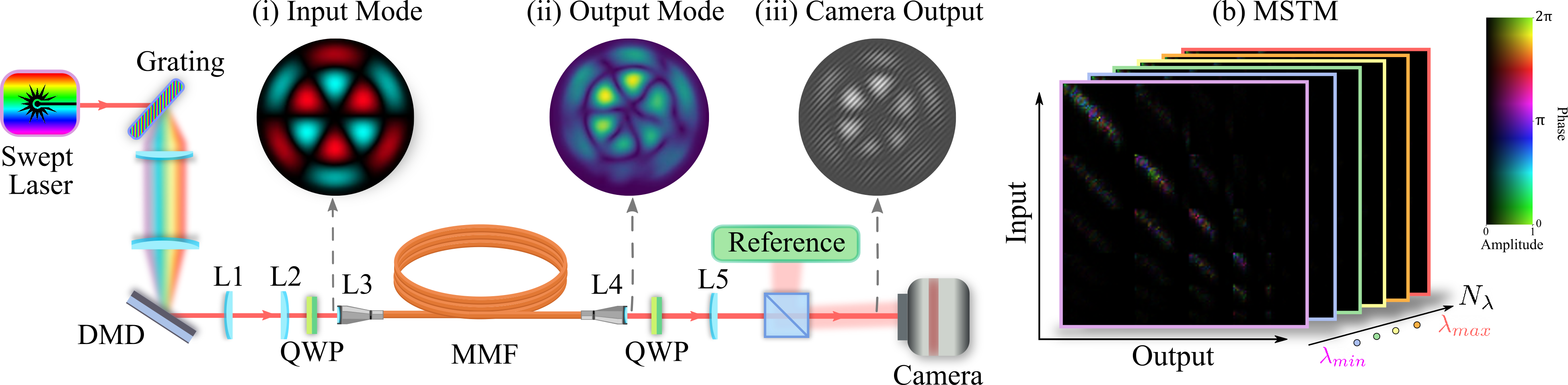}
\caption{\textbf{Procedure to measure the fiber MSTM.} A Swept laser source sets the wavelength and illuminates a DMD where holography is used to generate an input complex field (i) in the first diffraction order, which is launched into a 125-mode MMF. The output mode (ii) is interfered with a reference plane wave on the camera (iii), where off-axis holography is used to recover the output complex field. The MSTM can then be reconstructed by measuring all 125 modes for a number of wavelengths, depending on the desired temporal resolution and extent to be captured (lenses L1 $=250$ $mm$, L2 $=150$ $mm$, L3 $=7.5$ $mm$, L4 $=7.5$ $mm$, L5 $=500$ $mm$). }
\label{fig:MSTM_procedure}
\end{figure}

To obtain the transmission matrix (TM), it is crucial to have control of the input and output complex fields of the MMF (phase and amplitude). Fig.~\ref{fig:MSTM_procedure} depicts how we achieve this experimentally. For the input complex field, we use a digital micromirror device (DMD, Vialux V-650L NIR) as it features a high frame-per-second rate, which is well suited for scanning across many modes and multiple wavelengths. The cost of this speed is that the DMD pixels can only perform binary amplitude modulation (on or off) such that we cannot directly modulate the complex field. Instead we must employ holography methods. In this work we use the well-tested Lee holography method and encode our complex field pattern in the first-order diffraction from our DMD \cite{lee1979binary}. This allows us to sequentially generate complex fields at high speed but at the expense of large loss.

Ideally, using this procedure we can generate arbitrary complex fields. However, in practice DMD screens feature warping which imposes phase distortions on the modulated field. To accurately generate spatial modes we use an aberration correction algorithm which experimentally characterizes and counteracts the phase distortion by applying a correction pattern to the DMD screen. We use Zernike polynomials to optimize the point-spread function in the Fourier plane of the DMD with a lens and camera, following \cite{popoff2024practical}.

\medskip

At the input, we use a $1550$ $nm$ swept laser (Santec TSL-550) that is collimated and incident onto the DMD, allowing us to prepare arbitrary complex input modes for our MMF. The MMF used (Thorlabs FG050LGA) has a core diameter of $a =\SI{ 50 }{\mu m}$, NA of $0.22$, and a length of $40m$, for which we calculate 125 theoretical CP modes of the MMF at a wavelength of $1550$ $nm$ by solving the unified dispersion equation \cite{gloge1971weakly}. While experimentally it might seem that a more accurate TM could be achieved with a larger, oversampled spatial basis (plane wave modes, for example), we find that by analyzing the SVD decomposition of the TM, the CP basis captures the majority of the fiber's behavior while retaining the advantage of data compactness, which is important for speed and memory when measuring the full MSTM. Lastly, since the input wavelength is changed during measurement, the angle of the diffraction orders from the DMD also change. To compensate for this and ensure that we are well-aligned to the MMF's input facet for all wavelengths, we use a transmissive diffraction grating which is placed before the DMD along with a 4f-system made from appropriately chosen lenses ( see Fig.~\ref{fig:MSTM_procedure}).

\subsubsection{Complex field measurement}
We use an InGaAs camera (Hamamatsu C14041-10U) to measure the output modes from the MMF. This was chosen with the interest of measurement speed as it is capable of measuring images with a $\sim120$Hz refresh rate at $1550$ $nm$. While this was the fastest camera available to our knowledge, the frame rate is still 2 orders of magnitude smaller than the DMD and ultimately the bottleneck for measurement speed of the MSTM. Like most commercial cameras, we are limited to measuring the intensity of the light only, to retrieve the phase at the MMFs output facet, we use off-axis holography \cite{goodman1967digital}. This technique uses interference between the field emerging from the MMF---typically referred to as the object field---with a large, flat-phase reference field, producing an intensity-only interferogram on the camera, as illustrated in Fig.~\ref{fig:MSTM_procedure}. Using a mirror to tilt the reference beam's angle to the object beam, the period and angle of the interference fringes can be adjusted, which corresponds to shifting the position of the diffraction orders in the Fourier domain. Tuning the fringes correctly lets us separate the object field from the zero-order terms and thereafter digitally filter it with a simple mask and shift. Then, applying an inverse Fourier Transform we recover the full complex field of the object \cite{sanchez2014off}. In the experiment, we use a 90:10 fiber beamsplitter on our tuneable laser where the larger portion of the power is fed to the setup and the rest is expanded with a large collimator lens to use as the reference field. It is important to remark that path length matching the reference to the object is crucial to stabilize the phase drift of the interferogram on the camera as much as possible. However, even at perfect path lengths the phase between the reference and object beams, and therefore the fringes in the off-axis holography, typically feature some level of drift which can come from thermal expansion in the fiber, setup instability, or other macroscopic sources. This appears as phase noise along the spatial mode and wavelength axis of the MSTM. To avoid this, we track the drift using a reference spatial mode, which lets us interpolate and remove the drift from our data, reducing the phase noise of the MSTM. This tracking is done along the spatial mode axis and the wavelength axis of the MSTM.

\medskip

While the digital portion of the off-axis holography is algorithmically simple to employ in Python (using SciPy and NumPy), it is computationally slow for the large number of images we need to process. We found using Joel Carpenter's high-speed library "digHolo" \cite{carpenter2022digholo} to speed up the process tremendously and extend our gratitude.

\subsubsection{Retrieving the MSTM}

Using the swept laser to tune the wavelength, the DMD to generate complex fields and the off-axis holography to measure the fields, we can find the relation between input complex fields and their outputs and therefore measure and construct the MSTM as in Fig.~\ref{fig:MSTM_procedure}. We measure the MSTM of a $40m$ MMF, which allows for large modal dispersion, giving us more temporal extent to use for the high-dimensional measurements. The MSTMs used in this work were measured by launching all $125$ theoretical CP modes over 421 wavelengths such that the dimensions of the MSTMs were $421 \times 125 \times 125$. We used a wavelength resolution of $3.8$ $pm$ across a $1.6$ $nm$ range, which, when transformed into the TRTM, yields $5$ $ps$ temporal resolution over $1.65ns$. This was chosen to provide us enough temporal extent to eventually encode 11 time-bins, considering our SNSPD detectors have a jitter of about $40$ $ps$.

\subsection{Time-bin measurements}

\subsubsection{Constructing $\tau$-modes}
Once the MSTM is obtained, the TRTM can be computed by performing a Fourier transform along the spectral axis, as described in Eqs.~\ref{eqn:MSTM2}-\ref{eqn:TRTM}. We aim to find a basis of spatial modes ($\tau$-modes) for both the input and output spaces which maximize scattering to specific time delays $\tau^{(n)}$ and, furthermore, to ensure these are individually addressable, they must form an orthonormal basis. Rather than attempt to target arbitrary delays, we use the delays which the TRTM is best at scattering to . To this end we introduce Algorithm~\ref{alg:tau_algo}, in which we solve for the first $\tau$-mode by performing a singular value decomposition (SVD) of every time-slice, $ \tilde{T} (t) $, of the TRTM. By choosing the largest leading singular value (among all the $\tau$) and its respective left and right vectors/modes, we form the first '$\tau$' mode consisting of the field profiles for both the input $\tilde v^{(1)}$ and the output $\tilde u^{(1)}$. To search for subsequent '$\tau$' modes which are orthonormal to this first mode, we remove the presence of this mode from the TRTM by projecting the $\tau$-mode out using $\tilde{T}^{(n+1)}(t) = \Pi^{\perp}_{u} \tilde{T}^{(n)}(t) \Pi^{\perp}_{v}$, and then iterate the whole process on this updated TRTM for each '$\tau$' mode index $n$. We are then left with a set of orthogonal input and output $\tau$- modes with unique delays $\tau$. For ideal fibers this method would be closely related to principle modes as discussed in Section \ref{sec:tau_principal_modes}\cite{carpenter2017comparison}.

\begin{algorithm}[H] \label{alg:tau_algo}
\caption{$\tau$-mode construction}
\begin{algorithmic}[1]
\Inputs{$\tilde{T}(t)$}
\Initialize{$\tilde{T}^{(1)}(t) = \tilde{T}(t)$}
\For{$ n \in N $}
\State $\tilde{T}^{(n)}(t) \sim u_1(t) \sigma_1(t) v_1^{\dagger}(t) $
\Comment{leading singular values (SV) from SVD decompositions}
\State $ \tau^{(n)} = \argmax_t({\sigma_1(t)}) $
\State $\tilde{u}^{(n)} = u_1(\tau^{(n)}), \quad \tilde{v}^{(n)} = v_1(\tau^{(n)}) $
\Comment{Find delay and modes from largest SV}
\State $\Pi^{\perp}_{u}  = I - \ketbra{\tilde{u}^{(n)}}, \quad \Pi^{\perp}_{v} = I - \ketbra{\tilde{v}^{(n)}}$
\State $\tilde{T}^{(n+1)}(t) = \Pi^{\perp}_{u} \tilde{T}^{(n)}(t) \Pi^{\perp}_{v}$ \Comment{project $\tau$-mode out of current TRTM}
\EndFor
\State \Return $\tau^{(n)}, \quad \tilde{u}^{(n)}, \quad \tilde{v}^{(n)}  \, \forall \, n$
\end{algorithmic}
\end{algorithm}

\subsubsection{Spatial projections and SNSPD detection}
For detecting photons, we use superconducting nanowire single-photon detectors (SNSPDs, Quantum Opus) via SMF. However, at the MMFs output facet, the $\tau$-modes still have spatial mode structure which is incompatible with a SMF. We therefore employ an ``intensity-flattening" scheme \cite{bouchard2018measuring} which ensures high-quality spatial projective measurements of complex fields at the expense of higher loss. This is achieved with an SLM (Holoeye Pluto-2.1 LCOS), a 4f-system (f-300mm, f-150mm) and a coupling lens (f-7.5mm) for the SMF. Here, the incident complex field to be measured is "flattened" on the SLM which displays a hologram of its phase-conjugate structure. The back-propagated beam from the SMF is expanded by the 4f-system to be much larger than the incoming MMF field, such that its intensity distribution is approximately flat in the region of the incoming field. The incoming field can then be coupled to the SMF, with all spatial information "erased" and only the temporal profile of the tau mode remaining. The photon detection events are fed to the time tagger logic unit (Swabian Instruments) which are timed against a reference clock signal in the form of an RF signal coming from the pulse laser. In summary, we generate $\tau$-modes using a DMD and launch them into a multimode fiber (MMF). At the output, an SLM is used to manipulate these modes and coherently couple them into a single-mode fiber connected to an SNSPD detector. This approach allows us to program arbitrary transformations that exploit the full range of temporal delays supported by the MMF, enabling high-dimensional, "Franson-like" interferometric measurements.

\medskip

\begin{figure}[t!]
  \centering
  \includegraphics[width=\columnwidth]{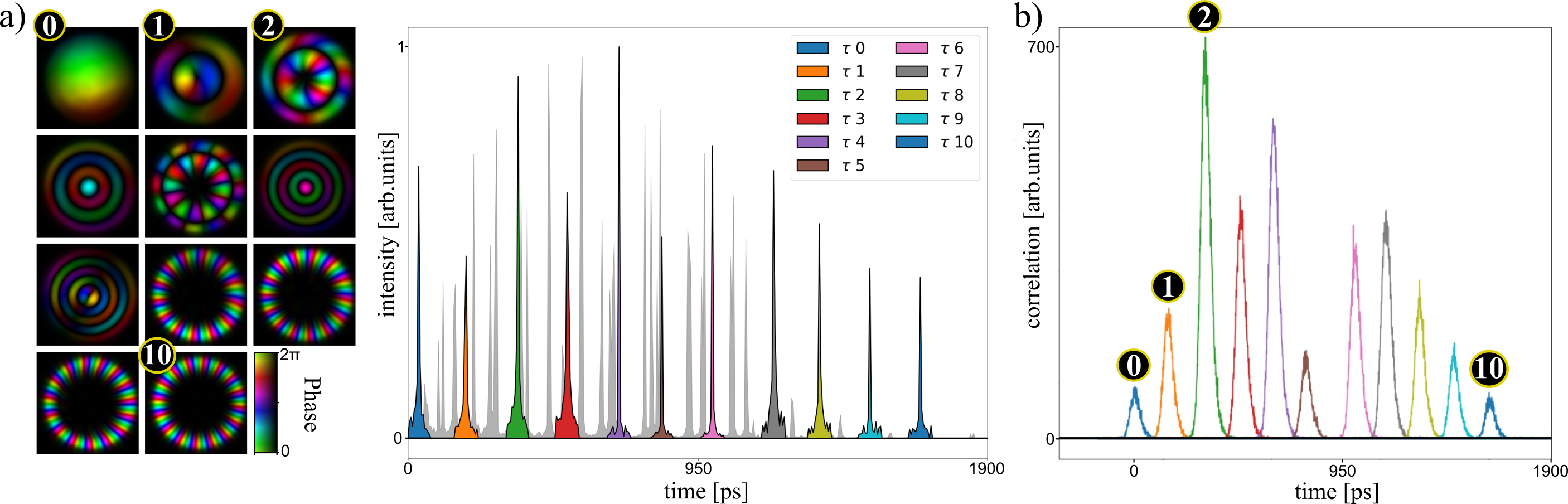}
  \caption{\textbf{Temporal measurements of $\tau$-modes} a) The input spatial patterns of 11 $\tau$-modes, $\tilde v^{(n)}$, along with their temporal profiles shown in the colored plots. The grayed out curve in the background is the temporal profile of the entire TRTM, found from the absolute value squared of the TRTM summed in its input and output spatial axes. b) We experimentally measure the $\tau$-modes using a pulsed laser and an SNSPD. The $\tau$-modes are sequentially displayed on the DMD and timed against a clock signal from the laser.
  }
  \label{fig:tau_times}
\end{figure}

To make temporal measurements, we swap the tunable laser with a pulsed laser (MENLO ELMO 1560) with $<150$ $fs$ pulse duration at $100MHz$ rep.rate and a bandwidth of $1560 \pm 30$ $nm$. We attenuate it to the appropriate single-photon level and additionally bandpass filter it to $1550.17 \pm 0.1$ $nm$ (increasing the pulse duration to about $30$ $ps$). The bandpass is used to ensure that the pulses fall within the $1.6$ $nm$ bandwidth of the MSTM. The holograms corresponding to $\tau$-modes or their superpositions can then be displayed on the DMD and launched into the MMF, where each $\tau$-mode will disperse and pick up its associated delay as can be seen in Fig.~\ref{fig:tau_times}.

\subsubsection{Loss budget}
We characterize the loss of the setup by measuring the transmission through each stage of the setup from the input SMF to the output SMF going to the SNSPD. These different stages can be seen in Fig.~\ref{fig:concept_interferometer} and \ref{fig:MSTM_procedure}. we measure the grating at the input to have $81.3\%$ transmission. The DMD generates $\tau$-modes with an efficiency of $1\%$ on average, which is primarily limited by the binary modulation of the DMD and loss due to mode shaping. Spatial-mode coupling to the MMF is about $10-20 \%$  efficient. After the MMF, we post-select one polarization, which leaves us with $66.5\%$ transmission. Lastly, when projecting the $\tau$-modes into the SMF via the SLM we observe an efficiency of approximately $8.2\%$ on average. It should be noted that in practice these efficiencies are spatial-mode-dependent and the mentioned efficiencies are averaged. The loss in our experiment can be mainly attributed to the DMD mode generation and SLM projection. This could be avoided by using an SLM instead of a DMD to generate modes and employing more efficient techniques for mode measurement.

\subsubsection{Franson interferometry}
\label{sec:SuppFranson}

\begin{table}[t!]
\caption{(Supplemental Material) Fidelity for the states $\ketbra{v_a^{\mu}}$ for dimensions $d=2,4,11$.}
\label{tab:S_fid_all_dims}
\begin{ruledtabular}
\begin{tabular}{lcccccccc}
 & \multicolumn{2}{c}{$d=2$}
 & \multicolumn{4}{c}{$d=4$}
 & \multicolumn{1}{c}{$d=11$} \\
\cline{2-3}\cline{4-7}\cline{8-8}
$a$
& $\mu=0$ & $\mu=1$
& $\mu=0$ & $\mu=1$ & $\mu=2$ & $\mu=3$
& $\mu=0$ \\
\hline
0
& $98.1 \pm 0.9\%$ & $97.8 \pm 0.9\%$

& $95.4 \pm 1.4\%$ & $96.0 \pm 1.4\%$ & $96.4 \pm 1.4\%$ & $96.1 \pm 1.4\%$

& $86.4 \pm 0.4\%$ \\
1
& $97.9 \pm 0.7\%$ & $98.8 \pm 0.7\%$

& $96.3 \pm 1.4\%$ & $96.6 \pm 1.4\%$ & $96.2 \pm 1.4\%$ & $96.3 \pm 1.3\%$

& $84.2 \pm 0.8\%$ \\
2
& \ldots & \ldots

& $96.3 \pm 1.6\%$ & $96.2 \pm 1.4\%$ & $96.5 \pm 1.4\%$ & $96.3 \pm 1.4\%$

& $86.3 \pm 0.8\%$ \\
3
& \ldots & \ldots

& $96.6 \pm 1.4\%$ & $96.3 \pm 1.3\%$ & $95.8 \pm 1.4\%$ & $96.5 \pm 1.4\%$

& $86.0 \pm 0.8\%$ \\
4
& \ldots & \ldots
& \ldots & \ldots & \ldots & \ldots
& $83.7 \pm 0.8\%$ \\
5
& \ldots & \ldots
& \ldots & \ldots & \ldots & \ldots
& $83.0 \pm 0.8\%$ \\
6
& \ldots & \ldots
& \ldots & \ldots & \ldots & \ldots
& $83.7 \pm 0.9\%$ \\
7
& \ldots & \ldots
& \ldots & \ldots & \ldots & \ldots
& $84.8 \pm 0.9\%$ \\
8
& \ldots & \ldots
& \ldots & \ldots & \ldots & \ldots
& $82.9 \pm 0.8\%$ \\
9
& \ldots & \ldots
& \ldots & \ldots & \ldots & \ldots
& $84.8 \pm 0.8\%$ \\
10
& \ldots & \ldots
& \ldots & \ldots & \ldots & \ldots
& $86.2 \pm 0.8\%$ \\
\end{tabular}
\end{ruledtabular}
\end{table}

\begin{figure}[t!]
  \centering
  \includegraphics[width=0.4\columnwidth]{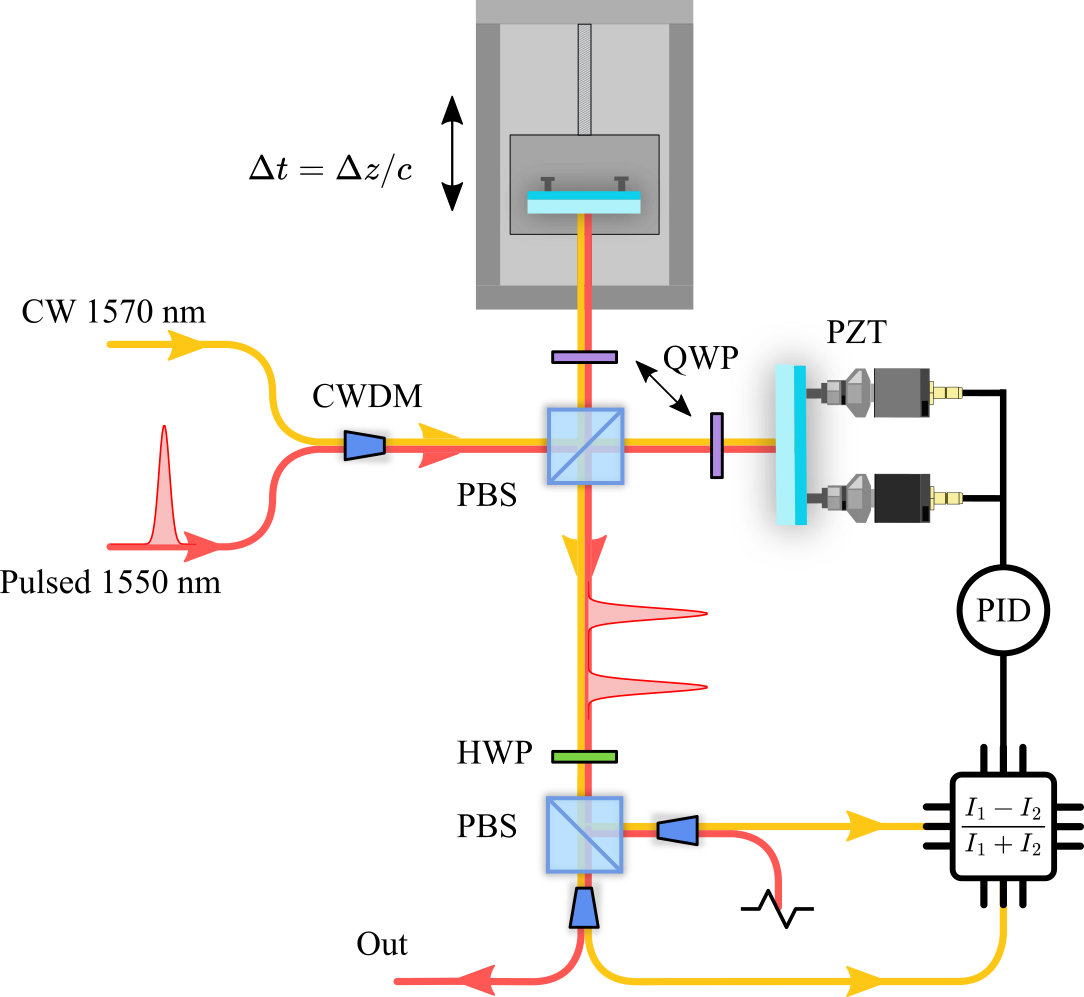}
  \caption{\textbf{Franson interferometer and phase-locking PID loop}. A $1570$ $nm$ CW locking laser is multiplexed with the $1550$ $nm$ pulsed signal using CWDMs. After demultiplexing at the output, the $1550$ $nm$ channel is sent to the DMD-MMF-SLM setup, while the $1570$ $nm$ light is detected by a reverse-biased photodiode and fed to a PID controller. The resulting feedback drives a piezo-mounted mirror to stabilize the interferometer phase, with a tunable setpoint enabling arbitrary phase preparation.
  }
  \label{fig:Franson}
\end{figure}

\begin{figure}[b!]
\includegraphics[width=0.8\columnwidth]{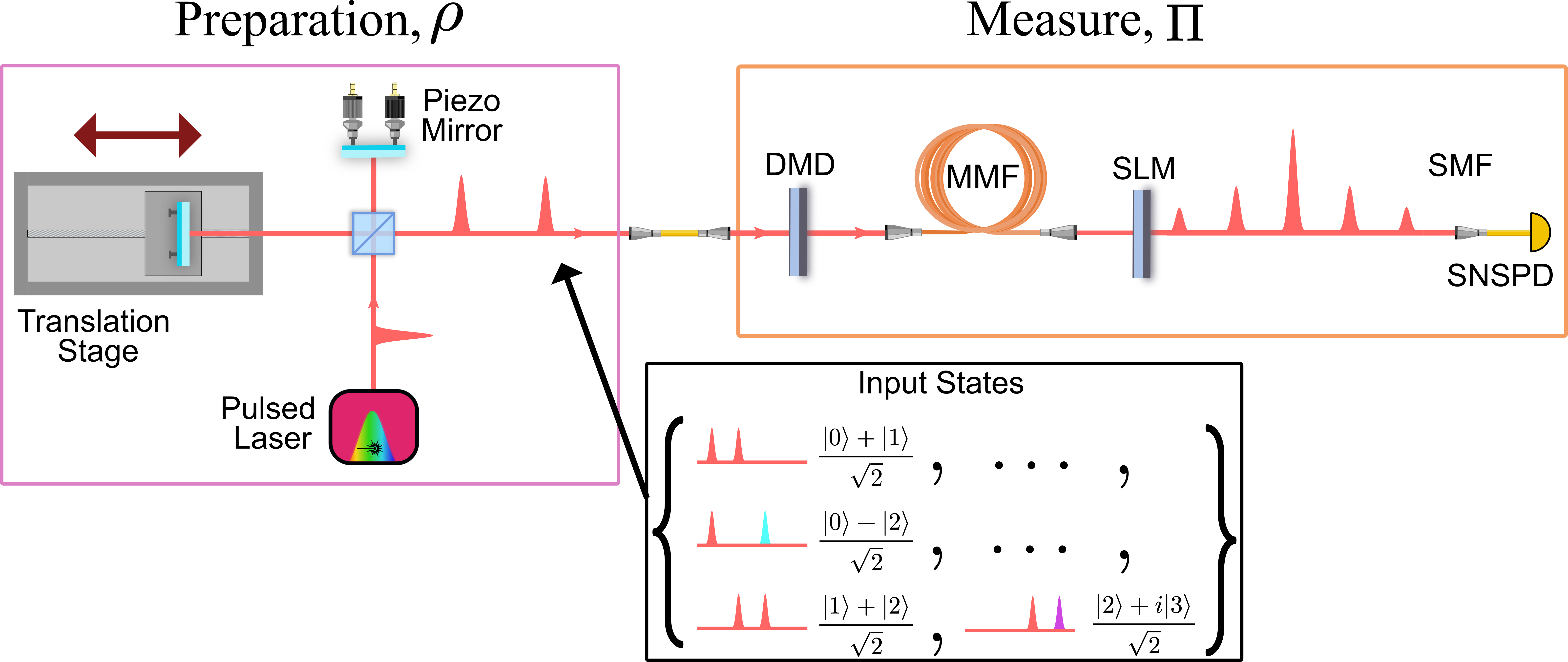}
\caption{\textbf{Prepare and measure setup.} The Franson interferometer on the left prepares a 2-dimensional time-bin state by varying the delay with a translation stage and setting (and locking) the phase with a piezo mirror. By preparing enough states in the 2-dim subspaces of the high-dimensional space of the DMD-MMF-SLM setup, we can perform quantum measurement tomography and quantify the fidelity of our setup as a measurement device.}\label{fig:prep_meas}
\end{figure}

In order to validate the results of our DMD-MMF-SLM setup, we perform a prepare-and-measure experiment. We do so by placing an unbalanced Michelson, also known as a Franson interferometer \cite{franson1989bell}, in front of our main setup for state preparation. This consists of a short arm and a long arm with a motorized translation stage (Standa 8MTL1401-300), allowing for a tunable delay. Phase drift naturally occurs in the setup as a result of environmental fluctuations; this drift is actively compensated using a locking laser system and a PID feedback loop. Here, a two channel ($1550$ $nm$, $1570$ $nm$) coarse wavelength division multiplexing (CWDM, OPNETI CWDM-1x2-1571-900-1-1-FC) fiber is used to multiplex the pulsed signal at $1550$ $nm$ with a CW locking laser at $1570$ $nm$, See Fig.~\ref{fig:Franson}. Additional CWDMs demultiplex the signal and locking laser at the output ports of the interferometer where InGaAs photodiodes (PDs, Thorlabs FGA015) detect the intensity of the locking laser which is now directly dependent on the phase between the interferometer arms. An analog divider circuit computes and outputs the difference-over-sum of the two PD voltages, which is the input signal for the PID controller. The PID controller consists of a Red Pitaya STEMlab 125-14 equipped with the PyRPL library \cite{neuhaus2017pyrpl}. The PID output signal is fed to a piezo-actuated mirror (Polaris $\textregistered$-K1S3P) on the short arm of the interferometer, which is used to actively adjust and lock the relative phase between the two arms to arbitrary setpoints between -180 to 180 degrees. This not only ensures phase stability but also allows preparation of arbitrary phases to perform measurement tomography. Four setpoints at $-135, -45, 45, 135$ degrees are used. The locking wavelength is then tuned at each stage position to prepare time bin states carrying $+,-,i,-i$ phases. The states are then fed to the main setup.

\subsubsection{Quantum tomography}

By moving the translation stage of the Franson interferometer and locking the phase at specific setpoints, we can prepare arbitrary 2-dimensional time-bin input states to probe our setup. However, since the device is high-dimensional, to tomograph it, we need to prepare enough linearly independent 2-dimensional states to span the high-dimensional space of the setup. This means we need to prepare all possible 2-dimensional subspace combinations of the $\tau$-modes. For instance, if the setup is performing 4-dimensional time-bin measurements, we need to prepare 6 different $\tau$-mode combinations with the Franson interferometer (which means moving the motorized stage to 6 different positions) all for which we measure 4 different phases. Completing these measurements for each mutually unbiased (MUB) element we can full tomograph our device as a projector, known as quantum measurement tomography. This was done for 4 dimensions as well as for 11-dimensions, there the number of stage positions required is 55 ($C(2,11)$). It should be noted that since the TRTM has a temporal resolution of $5$ $ps$ experimental fine-tuning of the stage positions had to be done to achieve maximal visibility for each $\tau$-mode combination.


\section{Additional Theory}

\subsection{MSTM and TRTM theory}

To better understand the TRTM and analyze the nature of the dispersion inside the MMF we employ the formalism of linear operators. 
We will see that a single Fourier transform to the $\omega$ axis of the multi-spectral transmission matrix (MSTM) then corresponds to the Time-Resolved Transmission Matrix (TRTM) whose $\Delta t=t-t'$ axis now corresponds to a time delay of $\Delta t$.
We can start by describing the MSTM as operator $\hat{T}$, acting on the input transverse field at position $x'$ and frequency $\omega$ and outputting to $x, \omega$. A general transformation may alter the frequency $\omega$ of the field however we consider linear transformations which act like,
\begin{equation} \label{eqn:MSTM2}
    \hat{T} = \int \diff^2x  \diff^2x' \diff \omega T(x,x';\omega)\ket{x,\omega}\bra{x',\omega}
\end{equation}

The inverse Fourier transform $\hat{F}$ between frequency $\omega$ and time $t$,

\begin{equation} \label{eqn:Fourier}
    \hat{F} = \int \diff t \diff\omega \, e^{i \omega \tau} \ket{t}\bra{\omega}
\end{equation}
allows to explore the TRTM by applying \ref{eqn:Fourier} to \ref{eqn:MSTM2} (at both the input and output spaces) to find:

\begin{align}
    \hat{\tilde{T}}  &= \hat{F} \hat{T} \hat{F}^{\dagger}   \\
     &= \int \diff^2x  \diff^2x' \diff t \diff t' \diff \omega \diff \omega'' \diff \omega''' T(x, x';\omega)  e^{i \omega'' t} e^{-i \omega''' t'} \ket{t}\bra{\omega''} 
 \ket{x,\omega}\bra{x',\omega}   \ket{\omega'''}\bra{t'} \\
    &= \int \diff^2x  \diff^2x' \diff t \diff t' \diff \omega T(x, x'; \omega) e^{i \omega (t -t')}\ket{x,t}\bra{x',t'} \\
    &=  \int \diff^2x \diff^2x' \diff t \diff t' \tilde{T}(x , x';t-t') \ket{x,t}\bra{x',t'} \label{eqn:TRTM}
\end{align}
Therefore the TRTM describes the time-delay $\Delta t=t-t'$.

\subsection{Transverse field bases}
Rather than dealing with the entire, continuous transverse field space we work with (undercomplete) bases chosen to approximately match the modes transmitted by our MMFs, thereby efficiently characterizing the fields and their transformations with fewest parameters (as discussed in App\ref{sec:complex_field_generation}). Given a set of orthonormal basis functions $\{e_i(x)\}_i$ corresponding to transverse fields, we define our basis modes,
\begin{align} \label{eqn:transverse_basis}
    \ket{e_i}&= \int \diff^2x \, e_i(x) \ket{x} \\   
\end{align}
Our MSTM and TRTM may then be defined with respect to these bases, for instance,
\begin{align}
    \hat{\tilde{T}}  &=   \sum_{ij} \int \diff t \diff t' \tilde{T}_{ij}( t- t') \ket{e_i, t}\bra{e_j, t'} \label{eqn:TRTM_mode_basis}
\end{align}
 and we will interchangeably use $\tilde T(\Delta  t)$ to refer to the matrix with elements $\tilde T_{ij}(\Delta  t)$. 

\subsection{Impulse response from tailored spatial modes}

We use the TRTM operator formalism to describe the device in terms of impulse response functions and show how that can be viewed as a time-bin (Franson) interferometer. If we prepare the $n$th input and output spatial $\tau$-modes: $\ket{v^{(n)}}=\sum_a v^{(n)}_a \ket{e_a}$ and $\ket{u^{(n)}}=\sum_b u^{(n)}_b \ket{e_b}$ the MMF will perform the temporal impulse response, $\hat{I}^{(n)}$:

\begin{align} \label{eqn:impulse_response}
    \hat{I}^{(n)} =\bra{ u^{(n)}} \hat{\tilde{T}} \ket{ v^{(n)} } &=   \sum_{ab} \int \diff t \diff t' \, u^{(n)*}_b  v^{(n)}_a \tilde{T}_{ab}( t- t') \ket{t}\bra{t'} \\
     &= \int \diff t \diff t' \, I^{(n)}(t-t') \ket{t}\bra{t'}
\end{align}
where the $I^{(n)}(t)$ are the impulse responses of the $\tau$-modes, which can be seen in Fig.~\ref{fig:tau_times}. Applying this to an arbitrary temporal input state $\ket{\Psi} = \int \diff t \, \phi(t)\ket{t}$ we find that the TRTMs operation can be described as a convolution with these impulse responses:

\begin{align} 
    \hat{I}^{(n)}\ket{\Psi}  &= \int \diff t \diff t' \diff t'' I^{(n)}(t-t') \phi(t'')\ket{t}\bra{t'} \ket{t''} \\
      &= \int \diff t \diff t' I^{(n)}(t-t') \phi(t)\ket{t} \\
      &= \int \diff t \left[I^{(n)} * \phi \right](t) \ket{t} \\ \label{eqn:tau_conv}
\end{align}
We would like to consider a family of time-bin modes which are localised about some time, $\tau$, with some strongly peaked temporal mode shape, $s(t)$, 
\begin{align*} 
    \ket{\tau}  &= \int \diff t \, s(t-\tau) \ket{t}
\end{align*}
Here $s(t)$ could be the temporal profile of our pump or signal laser pulse, for example. 
A set of times $\{\tau_a\}_a$, such that $|\braket{\tau_a}{\tau_b}|=\delta_{ab}$ would allow the modes $\{\ket{\tau_a}\}_a$ to form a good basis of time-bin modes. 

Additionally, we would like to choose $s(t)$ and $\{\tau_a\}_a$ such that our $\tau$-modes' impulse responses, $\hat{I}^{(n)}$, can elicit high-fidelity transformations between the chosen time-bin basis. Ie.
\begin{align*} 
    \hat{I}^{(n)}\ket{t}  \approx \ket{t -\tau_n}\\
\end{align*}
Ideally the impulse response in the convolution in eq.~\ref{eqn:tau_conv} will approximately act as a $\delta(t)$ distribution when applied to functions $s(t)$, and the delays $\tau^{(n)}$ should correspond to the delays between our chosen time-bin modes. This ideal behavior results in high-quality interference measurements between different time bins. Generally $s(t)$ should be chosen such that it is somewhat wider than the impulse responses, though it must necessarily by less wide than the spacings between the chosen times $\tau_a$ to maintain the orthogonality of the time-bin modes. To achieve this we chose a most-optimal subset of the available delays ($\tau^{(n)}$) to form our chosen time bin basis $\{\tau_a\}_a$ indexed by $a=1,...,d$ for $d=2, 4$ and $11$.  The impulse responses can then be considered to act like $\delta(t)$ distributions on the time-bin modes (as with the Franson interferometer).

\begin{align*} 
    \hat{I}^{(n)}\ket{\tau}  \approx \int \diff t \diff t' \, \delta(t - t' - \tau_n) s(t'-\tau) \ket{t'} = \ket{\tau -\tau_n}\\
\end{align*}
More generally, when the DMD displays a superposition of $\tau$-modes, $\sum_a f_a \ket{v^{(a)}}$, the resultant impulse response is a sum of the individual $\tau$-mode delays, determined by the measurement coefficients, $f_a$.

\begin{align*} 
    I^{\vec f}(t-t') =  \sum_{a}f_a I^{(a)}(t-t')  \approx \sum_{a}f_a \delta(t-t'-\tau_a)
\end{align*}
When considering the action of this operation on the modes of the time-bin basis, $\{ \ket{\tau_a} \}_a $, we obtain a discretized relation between the input and output time-bin modes, 

\begin{align*} 
    \hat{I}^{\vec f} &= \sum_k \int d\tau' I^{\vec f}(\tau_k- \tau'-\tau_i) \ket{\tau'}\bra{\tau_k} = \sum_{k,a} \int d\tau' f_a \delta(\tau_k - \tau'-\tau_a) \ket{\tau'}\bra{\tau_k} \\
    &= \sum_{k,a} f_a \ket{\tau_k - \tau_a}\bra{\tau_k}
\end{align*}
This leaves us with a simplified operator for our time-bin interferometer, which is identical to the idealized high-dimensional Franson interferometer.

Finally, for equidistant $\tau$-modes ($\tau_a = a \Delta$) we would have $\ket{\tau_k - \tau_a} = \ket{\tau_{k-a}}$ and thus
\begin{align*} 
    \hat{I}^{\vec f} &= \sum_{k,a} f_a \ket{\tau_{k -a}}\bra{\tau_k}
\end{align*}
admitting a neat matrix representation.
It is important to mention here that, while we aim to achieve this operator, due to the $\tau$-modes not being perfectly equally spaced or having perfect $\delta(t)$ distribution, we cannot achieve this exact operator, where experimentally we compensate for this with more measurements.

\subsection{$\tau$-modes and principal modes} \label{sec:tau_principal_modes}
Here we compare the $\tau$-modes that we construct to the well-established principal modes. If we start by assuming an ideal, straight, step-index multi-mode fiber, the solutions to the spatial modes come in the form of the circularly-polarized (CP) modes \cite{snyder1978modes,carpenter2017comparison, ploschner2015seeing}. For such an MMF the transmission matrix would act like a unitary matrix for each wavelength it supports and, assuming the same number of modes for each wavelength, can be represented in terms of the multi-spectral transmission matrix:

\begin{equation} \label{eqn:MSTM}
    T(\omega) = UD(\omega)U^{\dag}
\end{equation}

Here $U$ is a unitary matrix spanning the transverse spatial mode space of the MMF, $\omega$ the angular frequency and $D$ is a wavelength-dependent diagonal matrix. Since the CP modes have known dispersion we can rewrite this in the CP mode basis:

\begin{equation} \label{eqn:MSTM_LP}
    T(\omega) = \sum_{n=1}^{d} e^{i \beta_n(\omega) L} \ketbra{u_n}
\end{equation}

Here $n$ is the mode number, $\beta_n$ is the propagation constant of the $n$th mode and $u_n$ the spatial modes itself. The propagation constant can be expanded into:

\begin{equation} \label{eqn:propagation_constant}
    \beta_n(\omega) = \beta_n(\omega_0 +\Delta ) = \beta_n^{(0)} + \beta_n^{(1)}\Delta + \beta_n^{(2)}\Delta^2 + \beta_n^{(3)}\Delta^3 + ... + \beta_n^{(m)}\Delta^m
\end{equation}

We now calculate the first derivative of the MSTM:

\begin{equation} \label{eqn:Principal modes}
\begin{aligned}
     \pdv{}{\omega} T(\omega)\Bigr|_{\omega_0} &= \sum_{n=1}^{d} \pdv{}{\omega} e^{i \beta_n(\omega) L} \Bigr|_{\omega_0}  \ketbra{u_n} \\
      &= \sum_{n=1}^{d} L\pdv{\beta_n(\omega)}{\omega} \Bigr|_{\omega_0}  e^{i \beta_n(\omega) L}  \ketbra{u_n} \\
\end{aligned}
\end{equation}

From this we can construct the Wigner-Smith time delay operator, where the principal modes are obtained if one considers eq.~\ref{eqn:propagation_constant} only to the first-order \cite{carpenter2017comparison} (i.e. they have a flat wavelength dependence):

\begin{equation} \label{eqn:WS_operator}
\begin{aligned}
     T(\omega)^{\dag} \pdv{}{\omega} T(\omega)\Bigr|_{\omega_0} &= \sum_{n=1}^{d} L\pdv{\beta_n(\omega)}{\omega} \Bigr|_{\omega_0}  \ketbra{u_n} \\
     &= \sum_{n=1}^{d} L \beta_n^{(1)} \ketbra{u_n}
\end{aligned}
\end{equation}

For comparison, we instead approach the principal modes via Fourier transform of the MSTM yielding, the time-resolved transmission matrix (TRTM):

\begin{equation} 
    \tilde{T}(t) = \mathcal{F} \left\{ T(\omega) \right\} = \int e^{i \omega t} \sum_{n=1}^{d} e^{i \beta_n(\omega) L} \ketbra{u_n} dt
\end{equation}

If the propagation constant only contains the first-order expansion, then it is simple to show that we are left with:

\begin{equation} 
    \tilde{T}(t) = \sum_{n=1}^{d} e^{i \beta_0 L} \sqrt{2\pi} \delta \left( L\beta^{(1)}_n + t \right) \ketbra{u_n}
\end{equation}

In this ideal fiber case, these modes are the same as the principal modes from eq.~\ref{eqn:WS_operator}. The TRTM is not anymore unitary across its spatial axes, although globally, is still unitary. We can also consider second order dispersion such that we are instead left with:

\begin{equation}
    \tilde{T}(t) = \sum_{n=1}^{d} \frac{1}{\sqrt{-2 i  \beta^{(2)}_n}} e^{i \bigl( L\beta^{(0)}_n-\frac{(L\beta^{(1)}_n + t)^2}{4 L\beta^{(2)}_n} \bigr) }   \ketbra{u_n}
\end{equation}

To generalize this expression to the case of a real TRTM (with loss), we can convert the eigen-decomposition to a singular-value decomposition, by absorbing the phase term:

\begin{equation}
    \tilde{T}(t) = \sum_{n=1}^{d} \frac{1}{\sqrt{-2 i  \beta^{(2)}_n}} e^{i \bigl( -\frac{(L\beta^{(1)}_n + t)^2}{4 L\beta^{(2)}_n} \bigr) }   \ket{v_n}\bra{u_n}
\end{equation}

\subsection{Quantum tomography}
\label{sec:SuppTomo}
To fully characterise the behaviour of our MMF device as a quantum measurement, we perform tomography by probing the device with a tomographically complete set of input states and performing a constrained least-squares semi-definite program to invert the subsequent data to recover a description of the entire behavior of the device\cite{Hradil1997,qi2013,Smolin2012,flammia2012}. In our case, considering a $d$-dimensional time-bin basis for our input quantum states, the measurement our device (ie. DMD-MMF-SMF) performs is described by some positive $d$-dimensional operator, $\hat{\mathcal{O}}$, so that when an input state $\rho_n$ is incident, an SNSPD detector click in the correct time-bin occurs with probability (proportional to) $Y_n=\tr[\rho_n \hat{\mathcal{O}}] =: \mathcal{M}_n(\hat{\mathcal{O}})$. When the set of states $\{\rho_n\}_n$ is tomographically complete the system becomes invertible, in the sense that the measurement becomes uniquely identifiable via the inverse $\mathcal{M}^{-1}(Y) = \hat{\mathcal{O}}$. Alternatively, particularly in the presence of noise and/or less favorably structured sampling (unusual $\{\rho_n\}_n$ such as our 2-dimensional Franson preparations) it is advantageous to fit the data via a least-squares semi-definite program (SDP),
\begin{align}
\label{eqs:LSSDP}
\argmin_{\hat{\mathcal{O}}\geq0} |Y-\mathcal{M}(\hat{\mathcal{O}})|_F^2. 
\end{align}
In practise we have a pulsed laser source with a corresponding clock signal, using which we correlate time tags from the recieved photons and our clock, so that we can choose specific time-bins in the data to analyse. For instance, when the Franson is not present (or one arm is blocked), photons received in the time bins in our observed data correspond to having probed the measurement device with the computational basis, 
\begin{align}
\{\ket{n}\}_n \quad & \text{for } \quad 1\leq n \leq d \, .
\end{align}
When the Franson interferometer path lengths are appropriately chosen with their relative phase suitably locked, and with a prudent choice of time-bins to observe in the data, this corresponds to probing the measurement device with the $2$-dimensional subspace states, 
\begin{align}
\{\tfrac{1}{\sqrt{2}}\bigl(\ket{n} + \exp{i \phi_l} \ket{m}\bigr)\}_{nml} \quad &\text{for } \quad 1\leq n <m\leq d \quad , \quad \phi_l\in\{0,\pi/2,\pi,3\pi/2\}
\end{align}
This set is somewhat overcomplete owing to the $l=\pi$ and $l=3\pi/2$ being linearly dependent on the other measurements, which serves to determine the total count rate from our laser source. These states constitute a tomographically (over) complete set for tomography, with which the SDP Eqs.~\ref{eqs:LSSDP} can be performed. Having fit Eqs.~\ref{eqs:LSSDP}, we choose to normalise the recovered measurements by their trace, i.e.~$\hat{\mathcal{O}}:=\hat{\mathcal{O}}^*/\tr[\hat{\mathcal{O}}^*]$ and use the quantum state fidelity between the reconstructed $\hat{\mathcal{O}}$ and target $\hat \Pi$ measurement operators, since the target is rank 1, and projective/unit trace. This method allows us to focus on the quality of measurement rather than the specific efficiency which depends on many other factors such as our SNSPD efficiencies and full characterisation of our incident pulsed source power. Furthermore, since we are tomographing single elements from our measurement, one at a time, as many effective single-outcome measurements, there are not the completeness requirements for an entire POVM which would require quantum measurement tomography constraints (though there are plans to extend these measurements to multi-outcome counterparts in the future). Finally, to capture the effects of errors arising from imperfect measurements/data acquisition, we have characterised the phase noise in our Franson interferometer by repeating a standard 2-dimensional measurement at each locking point, $200$ times and extracted the standard deviation of the phase error. This error allows us to numerically obtain error bounds on our tomography by performing Monte-Carlo sampling consisting of simulating data (200 points per state) acquired with the characterised phase error and performing the tomographic inversion to obtain a distribution of reconstructed states and corresponding fidelities with which we cite the standard deviation in the reported values.

\end{document}